\documentclass{article}
\usepackage[top=1in, bottom=1.5in, left=1in, right=1in]{geometry}
\usepackage{graphicx} 
\usepackage{algorithm}
\usepackage{algpseudocode}
\usepackage{amsfonts,amsmath,amssymb,mathrsfs,dsfont,amsthm}
\usepackage{color}
\usepackage{natbib}
\usepackage{enumerate}
\usepackage{bbm,bm}
\usepackage{float}
\RequirePackage[colorlinks,citecolor=blue,urlcolor=blue]{hyperref}
\usepackage{multirow}

\newcommand{\Small}[1]{\textstyle #1 \displaystyle}
\newcommand{\ddr}{\mathrm{d}}

\newcommand{\comillas}[1]{``\,#1\,''}
\definecolor{iblue}{rgb}{0.1,0,0.75}
\definecolor{ired}{rgb}{0.9,0,0.1}
\definecolor{gray}{rgb}{0.5, 0.5, 0.5}

\newcommand{\Pmodal}{\hat{P}}

\newcommand{\Indic}{\mathds{1}}

\newcommand{\bbP}{{p}}

\newcommand{\sbm}{stochastic block model }

\newenvironment{Bluenote}[1]
{\begin{footnotesize}\par{\color{iblue}{#1}}}
{\par\end{footnotesize}}

\newif\ifbox
\boxtrue  

\title{Bayesian nonparametric community detection in assortative stochastic block models}

\author{Martina Amongero$^{1}$, Pierpaolo De Blasi$^{1,2}$\thanks{corresponding author \href{mailto:pierpaolo.deblasi@unito.it}{pierpaolo.deblasi@unito.it}}  \\[4pt]
        \small $^{1}$ESOMAS Department, University of Torino, Italy \\
        \small $^{2}$Collegio Carlo Alberto, Torino, Italy\\
}

\date{}

\providecommand{\keywords}[1]{\textbf{Keywords}: #1}

\begin{document}
\maketitle

\begin{abstract}
Structured data in the form of networks are increasingly common in a number of fields, including the social sciences, biology, physics, computer science, and many others. A key task in network analysis is community detection, which typically consists of dividing the nodes into groups such that nodes within a group are strongly connected, while connections between groups are relatively scarce. A generative model well suited for the formation of such communities is the assortative stochastic block model (SBM), which prescribes a higher probability of a connection between nodes belonging to the same block rather than to different blocks. A recent line of work has utilized Bayesian nonparametric methods to recover communities in the SBM by placing a prior distribution on the number of blocks and estimating block assignments via collapsed Gibbs samplers. However, efficiently incorporating the assortativity constraint through the prior remains an open problem. In this work, we address this gap by studying the effect of enforcing assortativity on Bayesian community detection and  identifying the scenarios in which it pays dividends in comparison with standard SBM. We illustrate our findings through an extensive simulation study.
\end{abstract}


\section{Introduction}

Available data in the form of networks has gained increasing attention in modern research, with applications ranging from biology (e.g., brain activity, protein interaction networks), to medicine (e.g., evolution of infectious disease), and economics (e.g., social networks and advertising policies). 
The study of community structures of networks has been a popular topic in Statistics. We speak of {\it assortative} communities \citep{New:10} when there are more connections within groups than between them. There exist several approaches to finding communities, either heuristic or model-based. The first category includes hierarchical clustering \citep{newman:2004}, spectral clustering \citep{Rohe:2011,Lei:Rin:15}, and maximization of modularity \citep{New:Gir:04,Blo:etal:08,Zha:Lev:Zhu:11}. 
In model-based approaches, rather than simple features of the network structure, communities are an intrinsic part of the generating process: the probability of nodes being connected or not depends on them being part of the same community or different communities. The most popular probabilistic model used to represent networks is the \sbm (SBM) of \citet{Hol:Las:Lei:83}. In its simplest form, it comprises a $k\times k$ symmetric matrix $P=(P_{j\,l})$ of probabilities of connection between nodes of equal ($j=l$) or different blocks ($j\neq l$), and a $k$-vector $\pi=(\pi_j)$ of probabilities that a node is assigned to a given block thus determining how much balanced -- i.e. similar in size -- the blocks are.

SBM has been extensively studied in terms of theoretical guarantees for community recovery  such as detection thresholds \citep{Mos:Nee:Sly:15,Mos:Nee:Sly:16,Abb:Ban:Hal:16} and minimax rates \citep{Zha:Zho:16,Gao:Ma:21}. These guarantees can be expressed in terms of the number of nodes and edges, together with the number of blocks $k$, the balance (or imbalance) of blocks, and the gap between the smallest within-block probability, $p=\min_{j}P_{j\,j}$, and the largest between-block probability, $q=\max_{j< l}P_{j\,l}$. We refer to the model as \textit{assortative} SBM whenever $p>q$ \citep{Ami:Lev:18}, a condition that makes blocks consistent with communities in the assortative sense.

Communities can be recovered via maximum likelihood estimation \citep{Ami:etal:13,Bic:Che:09}, 
semi-definite programming relaxations \citep{Ami:Lev:18}, Bayesian inference \citep{Sni:Now:97,Pas:Vaa:18}, and mean field variational methods \citep{Bic:Cho:Cha:Zha:13,Zha:Zho:20}.  
The number of blocks $k$ is of pivotal importance -- as in more traditional clustering applications -- in that it determines the intrinsic dimension of the model. Various approaches have been proposed to select $k$ such as modularity maximization \citep{Zha:Lev:Zhu:11}, the integrated complete data log-likelihood \citep{Com:Lat:15}, likelihood ratio tests \citep{Wan:Bic:17}, information criteria  \citep{Sal:Yu:Fen:17}, variational methods \citep{Lat:Bir:Amb:12,Zha:Yan:24}.
Recently, there has been a surge of interest in Bayesian community detection that incorporates the choice of $k$ by placing a prior distribution on it. Block assignments are then estimated via collapsed Gibbs samplers  \citep{mcdaid:murphy:friel, Gen:Bha:Pat:19, Leg:Rig:Dur:Dun:22}. 
%
Attempts to incorporate assortativity in Bayesian inference are scarce. In the case of fixed $k$, we refer to \citet{Pen:Car:16, Grib:etal:21} for the degree-corrected SBM \citep{Kar:New:11} and \citet{Gop:etal:12} for mixed-membership SBM 
\citep{Air:etal:08}. To the best of our knowledge, the only attempt in the unknown-$k$ setting is the technical report by \citet{Jia:Tok:21}. Enforcing the assortativity constraint $q<p$ prevents the use of independent and conjugate priors on connection probabilities. MCMC methods heavily rely on marginalization over model parameters, so a trade-off between algorithmic efficiency and flexibility in model specification arises, especially in the context of collapsed Gibbs samplers with $k$ unknown.
%

In this paper we propose a probabilistic framework for simultaneous estimation of the number of communities $k$ and the community memberships in assortative SBMs. 
The two main contributions are a novel prior specification on connection probabilities that enforces the assortativity constraint $q<p$, and an efficient Gibbs sampler for community detection when $k$ is unknown. We use synthetic networks to investigate the advantages of assortative SBM in community detection. In particular we focus on recovery regimes in which posterior inference benefits from imposing assortativity through the prior, accounting for the computation overhead due to the lack of conjugacy. 


The rest of the paper is organized as follows. In Section \ref{section:2}, we introduce the SBM, and discuss how to incorporate the assortativity constraint in the prior specification of the probability matrix $P$. In Section \ref{section:3}, we show how to extend existing algorithms to assortative modeling, building on the work of \citet{Nea:00} and \citet{Gen:Bha:Pat:19}. Finally, in Section \ref{section:simulation} we present a simulation study with networks generated from benchmark graphs that account for heterogeneity in node degree distributions and community sizes \citep{Lan:For:Rad:08}. Conclusions are presented in Section \ref{section:discussion}.


\section{Bayesian stochastic block modeling}\label{section:2}

In this section we introduce the \sbm and Bayesian community detection with $k$ fixed. Then we discuss how to include assortativity through the prior and illustrate its importance on a synthetic network.

Consider an undirected graph with no self-loops made of $n$ nodes, labeled $1,\ldots,n$, described by a symmetric $n\times n$ adjacency matrix $A$, with entry $A_{ij}=1$ if node $i$ and node $j$ are connected and $A_{ij}=0$ otherwise. Each node belongs to one of $k$ mutually exclusive groups, called blocks, labelled $1,\ldots,k$, that share a similar connectivity patterns according to
  $$\bbP(A_{ij}=1)=P_{z_i z_j}$$
independent across $i\neq j$. Here $P=(P_{ab})$ is the connectivity matrix, a $k\times k$ symmetric matrix with $P_{ab}\in(0,1)$ being the probability of an edge between nodes of blocks labelled $a$ and $b$. Also  $z_i\in \{1,\ldots,k\}$ is the block assignment, or label, of the $i$th node. We will refer to $z=(z_1,\ldots,z_n)$ as the vector of block assignments.
Finally, let $\pi=(\pi_1,\ldots,\pi_k)$ be probabilities of community assignments, so that $\bbP(z_i=a)=\pi_a$ for $a=1,\ldots,k$ independently across $i=1,\ldots,n$ . In summary, 
\begin{equation*}
\begin{aligned}
  z_i|\pi&\overset{\text{iid}}{\sim}
  \text{Multinomial}(1,\pi),\quad i=1,\ldots,n\\
  A_{ij}|z,P&\overset{\text{ind}}{\sim}
  \text{Bernoulli}(P_{z_i,z_j}),\quad i< j
\end{aligned}
\end{equation*}
The complete likelihood is given by
\begin{equation}\label{eq:likelihood_sbm}
\begin{aligned}
  \bbP(A,z|P,\pi)
  =\prod_{i<j}P_{z_i,z_j}^{A_{ij}}(1-P_{z_i,z_j})^{1-A_{ij}}
  \prod_i \pi_{z_i}
  &=\prod_{a\leq b}P_{ab}^{O_{ab}(z)}
  (1-P_{ab})^{n_{ab}(z)-O_{ab}(z)}
  \prod_a \pi_a^{n_a(z)}
\end{aligned}
\end{equation}
where
  $n_a(z)=\sum_{i=1}^n\Indic(z_i=a)$
is the number of nodes labelled $a$,
\begin{equation*}
  O_{ab}(z)=(\Small{\frac12})^{\Indic(a=b)}\sum_{i\neq j}\Indic(z_i=a,z_j=b)A_{ij},\quad
  n_{ab}(z)=(\Small{\frac12})^{\Indic(a=b)}\sum_{i\neq j}\Indic(z_i=a,z_j=b),
\end{equation*}
are the number of edges between nodes labelled $a$ and $b$ by the labelling $z$, and the maximum number of edges that can be created between nodes labeled $a$ and $b$, respectively. 


We complete the Bayesian model by specifying a prior on $\pi$ and $P$. The prior on $\pi$ is symmetric Dirichlet
\begin{equation}\label{eq:prior_pi}
     \pi \sim\text{Dirichlet}(\gamma,\ldots,\gamma),  
\end{equation}
for $\gamma>0$. The \textit{standard} prior specification on $P$ is to have $P_{ab}$ beta distributed and independent across $a$ and $b$: 
\begin{equation}\label{eq:prior_Pab}
    P_{ab}\overset{\text{iid}}{\sim}
  \text{beta}(\alpha,\beta), \quad  0\leq a\leq b\leq k
\end{equation}
for $\alpha,\beta>0$. The convenience is that one can exploit the standard beta-binomial update to derive the conditional of $P$ given $z$ and $A$:
\begin{equation}\label{eq:full_Pab}
  \begin{aligned}
   P_{ab}| A,  z &\overset{\text{ind}}{\sim}
  \text{beta}(O_{ab}(z)+\alpha,n_{ab}(z)-O_{ab}(z)+\beta).
  \end{aligned}
\end{equation}
We then sample from the conditional distribution of $z$ given $A$ and $P$ by integrating out the probabilities $\pi$.
According to \eqref{eq:prior_pi}, the conjugacy between the multinomial and Dirichlet distributions gives the marginal probability of label assignments
\begin{equation}\label{eq:p_z}
  \bbP(z)
  =
  \frac{\Gamma(\gamma k)}{\Gamma(\gamma)^k\Gamma(n+\gamma k)}\prod_{a} \Gamma(n_a(z)+\gamma)
\end{equation}  
from which one obtains the {\it prior predictive} probabilities
\begin{equation}\label{eq:prior_predictive}
  \bbP(z_i=a|z_{-i})=\frac{n_{a}(z_{-i})+\gamma}{n-1+\gamma k}
\end{equation}
where $z_{-i}$ is the vector of $n-1$ assignments but the $i$th one. The full conditional of $z_i$ is
\begin{align}
  \bbP(z_i=a|z_{-i}, A,P)
  &= \frac{\bbP(z_i=a,A|z_{-i}, P)}{\bbP(A|z_{-i}, P)}
  =\bbP(z_i=a|z_{-i})\frac{\bbP(A|z_i=a,z_{-i}, P)}{\bbP(A|z_{-i}, P)}\nonumber\\
  &= \bbP(z_i=a|z_{-i})\frac{\bbP(A|z_i=a,z_{-i}, P)}{\bbP(A_{-i}|z_{-i}, P)
  \bbP(A_i|A_{-i},z_{-i}, P)}\nonumber\\
  \label{eq:full_z}
  &\propto \bbP(z_i=a|z_{-i})
  \frac{\bbP(A|z_i=a,z_{-i}, P)}{\bbP(A_{-i}|z_{-i}, P)}
\end{align}
that is proportional to the product of the prior predictive probability and the {\it likelihood contribution to $z_i=a$}:
\begin{equation}\label{eq:like_contr}
  \frac{\bbP(A|z_i=a,z_{-i}, P)}{\bbP(A_{-i}|z_{-i}, P)}
  =\prod_{j>i} P_{a\,z_j}^{A_{ij}}
  (1-P_{a\,z_j})^{(1-A_{ij})}
  \prod_{j<i} P_{z_j\,a}^{A_{ji}}(1-P_{z_j\,a})^{(1-A_{ji})}
  =\prod_{j\neq i} P_{a\,z_j}^{A_{ij}}
  (1-P_{a\,z_j})^{(1-A_{ij})}
\end{equation}
Here $A_i$ refers to the connections involving the $i$th node and $A_{-i}$ to the $(n-1)\times(n-1)$ adjacency matrix upon removal of the $i$th node. 
We report the pseudo code of the Gibbs sampling algorithm next.

\begin{algorithm}[ht]
\caption{Gibbs sampler for SBM, k fixed}\label{alg:SBM_k}
\begin{algorithmic}
\Procedure{SBM-k}{}
\State{Require $n\times n$ adjacency matrix $A$, number of communities $k$, number of iterations $M$,} 
\State{prior hyperparameters $\gamma,\alpha,\beta$.}
\smallskip
\State {Initialize $z=(z_1,...,z_n)$ 
}
\smallskip
\For{each iter $j=1$ to $M$} 
    \State{- update $ P=(P_{ab})$ conditional on $z$ and $A$ from \eqref{eq:full_Pab} } 
    \For{each iter $i=1$ to n} 
    \State{- update $z_i$ conditional on $z_{-i}$, $P$ and $A$ from \eqref{eq:full_z} via \eqref{eq:prior_predictive} and \eqref{eq:like_contr}} 
    \EndFor
\EndFor
\EndProcedure
\end{algorithmic}
\end{algorithm}


\subsection{Assortative \sbm} \label{subsection:assortative_SBM}

\begin{figure}[ht]
    \centering
    \includegraphics[width=1\linewidth]{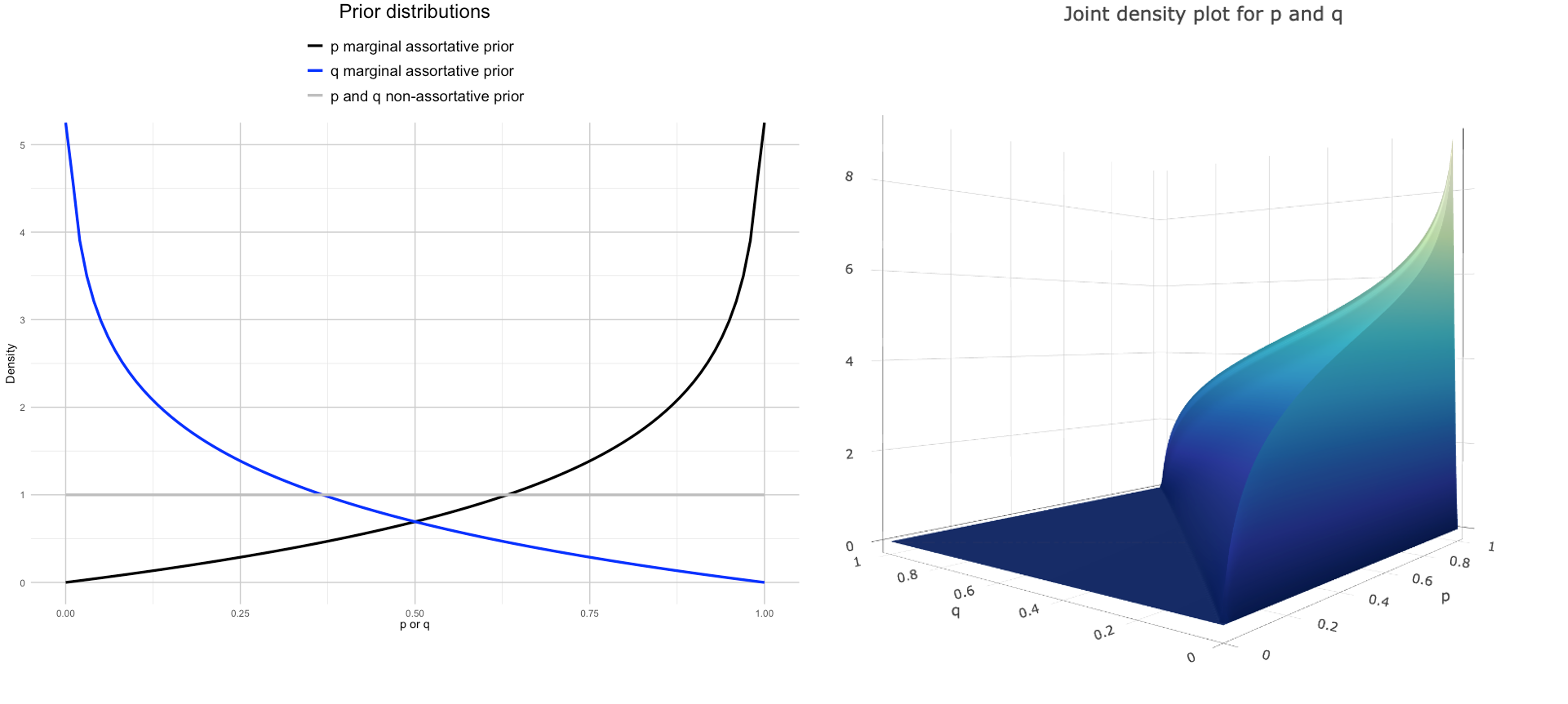}
\caption{
The left panel illustrates the marginal of $p$ and $q$, the grey horizontal line marks the uniform density. The right panel shows the joint density which is supported on $q<p$.
}
    \label{fig:prior}
\end{figure}

There exist two different notions of assortativity, cf. \citet[Definition 4.1]{Ami:Lev:18}. \textit{Strong} assortativity requires that any within probability is larger than the maximum between probability:
\begin{equation}\label{eq:strong_ass}
  \max_{a\neq b} P_{ab}<\min_a P_{aa},
\end{equation}  
{\it Weak} assortativity relaxes the above by requiring a diagonal dominance of the matrix $P$:
\begin{equation}\label{eq:weak_ass}
  q^*_a<P_{aa} \quad\text{for all }a,
\end{equation}
where
  $  q^*_a=\max_{b=a, c\neq a} P_{bc}
  =\max_{b\neq a, c= a} P_{bc}$.
%
We discuss first how to induce a stochastic order $q<p$ between two random probabilities $p,q\in[0,1]$. A 
computationally convenient scheme consists in  introducing an auxiliary variable $\epsilon\in[0,1]$, referred to as the {\it cutoff}, and set both $p$ and $q$ conditional on $\epsilon$ as truncated on $(\epsilon,1)$ and $(0,\epsilon)$, respectively. For illustration purposes we confine to $\epsilon$ uniformly distributed on $(0,1)$, $\epsilon\sim \text{U}(0, 1)$, extensions to $\text{beta}(\alpha,\beta)$ distribution can be worked out similarly. We have
\begin{gather*}
    \epsilon\sim U(0, 1), \\
    p|\epsilon \sim U(\epsilon, 1), \quad
    q|\epsilon \sim U(0, \epsilon).
\end{gather*}
The marginal and joint densities are then found to be:
\begin{gather*}
  f(p)= -\log(1 - p),\quad
  f(q)= -\log q \\
  f(p,q)=\int_q^p x^{-1} (1-x)^{-1}\ddr x
  =\log\frac{p(1-q)}{q(1-p)}\Indic_{\{q<p\}}
\end{gather*}
Figure \ref{fig:prior} visually represents these densities. 

Henceforth we analyze strong assortativity  \eqref{eq:strong_ass} and discuss how to implement weak assortativity \eqref{eq:weak_ass} in Section \ref{section:discussion}. We have in place of \eqref{eq:prior_Pab},
\begin{subequations}
\label{eq:prior_P_ass}
\begin{align}   
  \epsilon &\sim U(0,1),\\
  P_{aa}|\epsilon & \overset{\text{iid}} \sim  \text{U}(\epsilon,1), \quad  a=1,\ldots, k,\\
  P_{ab}|\epsilon & \overset{\text{iid}} \sim  \text{U}(0,\epsilon), \quad  1\leq a < b\leq k.
\end{align}
\end{subequations}
%
As for the conditional joint distribution of $P$ and $\epsilon$ given $A$ and $z$, we obtain  \begin{multline*}
  \{ P_{aa}\}_a, \{ P_{ab}\}_{a< b},\epsilon|A,z \propto
  \epsilon^{-k(k-1)/2} (1-\epsilon)^{-k}\\
  \prod_{a} P_{aa}^{O_{aa}(z)}(1-P_{aa})^{n_{aa}(z)-O_{aa}(z)}
  \Indic_{\{P_{aa}>\epsilon\}}  
  \prod_{a<b} P_{ab}^{O_{ab}(z)}(1-P_{ab})^{n_{ab}(z)-O_{ab}(z)} 
  \Indic_{\{P_{ab}<\epsilon\}}.
\end{multline*}
which leads to the following update for $P$:
\begin{equation}\label{eq:full_Pab_ass}
  \begin{aligned}
            P_{aa}| \epsilon, A, z &
            \sim
            \text{beta}_{(\epsilon,\ 1)}(O_{aa}(z)+1,n_{aa}(z)-O_{aa}(z)+1), \quad  1 \leq a\leq k,\\
            P_{ab}| \epsilon, A, z &
            \sim
            \text{beta}_{(0,\ \epsilon)}(O_{ab}(z)+1,n_{ab}(z)-O_{ab}(z)+1), \quad  1\leq a<b\leq k.
        \end{aligned}
\end{equation}  
To sample from truncated beta distributions $\text{beta}_{(\epsilon,\ 1)}(\cdot)$ and $\text{beta}_{(0,\ \epsilon)}(\cdot)$, one can use numerical approximation of the incomplete beta function or use data augmentation as in \citet{Dam:Wal:01}. As for the cut off $\epsilon$, we have
\begin{equation}\label{eq:full_epsilon}
  \epsilon|P, A, z  \propto
  \epsilon^{-k(k-1)/2} (1-\epsilon)^{-k}
  \Indic_{  \{q<\epsilon<p\}},\quad
  p=\min_{a}P_{aa},\ q=\max_{a< b}P_{ab}.
\end{equation}
Taking inspiration again from \citet{Dam:Wal:01}, we can sample from \eqref{eq:full_epsilon} by introducing a latent variable $y$ which has joint density with $\epsilon$ given by
  $$\epsilon,y|P
  \propto \epsilon^{-k(k-1)/2}
  \Indic_{\{y<(1-\epsilon)^{-k},\ q<\epsilon<p\}}$$
The full conditional for $y$ is given by the uniform density on the interval $(0, (1-\epsilon)^{-k})$. As for $\epsilon$, the full conditional
  $$
  \epsilon|y,P\propto\epsilon^{-k(k-1)/2}
  \Indic_{\{\max\{q,1-y^{-1/k}\}<\epsilon<p\}}
  $$
can be easily sampled using the inverse cdf technique.
We report the Gibbs sampling algorithm for assortative-SBM next.

\begin{algorithm}[ht]
\caption{Gibbs sampler for assortative-SBM, k fixed}\label{alg:SBM_k_ass}
\begin{algorithmic}
\Procedure{a-SBM-k}{}
\State{Require $n\times n$ adjacency matrix $A$, number of communities $k$, number of iterations M,} 
\State{prior hyperparameters $\gamma,\alpha,\beta$.}
\smallskip
\State {Initialize $z=(z_1,...,z_n)$ and $\epsilon$}
\smallskip
\For{each iter $j=1$ to $M$} 
    \State{- update $ P=(P_{ab})$ conditional on $\epsilon$, $z$ and $A$ from \eqref{eq:full_Pab_ass} } 
    \State{- update $\epsilon$ conditional on $P$, $z$ and $A$ from \eqref{eq:full_epsilon} } 
    \For{each iter $i=1$ to n} 
    \State{- update $z_i$ conditional on $z_{-i}$, $P$ and $A$ from \eqref{eq:full_z} via \eqref{eq:prior_predictive} and \eqref{eq:like_contr}} 
    \EndFor
\EndFor
\EndProcedure
\end{algorithmic}
\end{algorithm}


\subsection{Illustrative example} \label{subsection:motivating-example-section}
We consider a synthetic network generated according to a SBM and intentionally designed to highlight the effect of enforcing the assortativity constraint in Bayesian community detection. Specifically, the example demonstrates that without enforcing assortativity the posterior can fail to recover the communities. We work with a fixed and exactly specified number of communities. 

\begin{figure}[ht]
    \centering
    \includegraphics[width=0.8\linewidth]{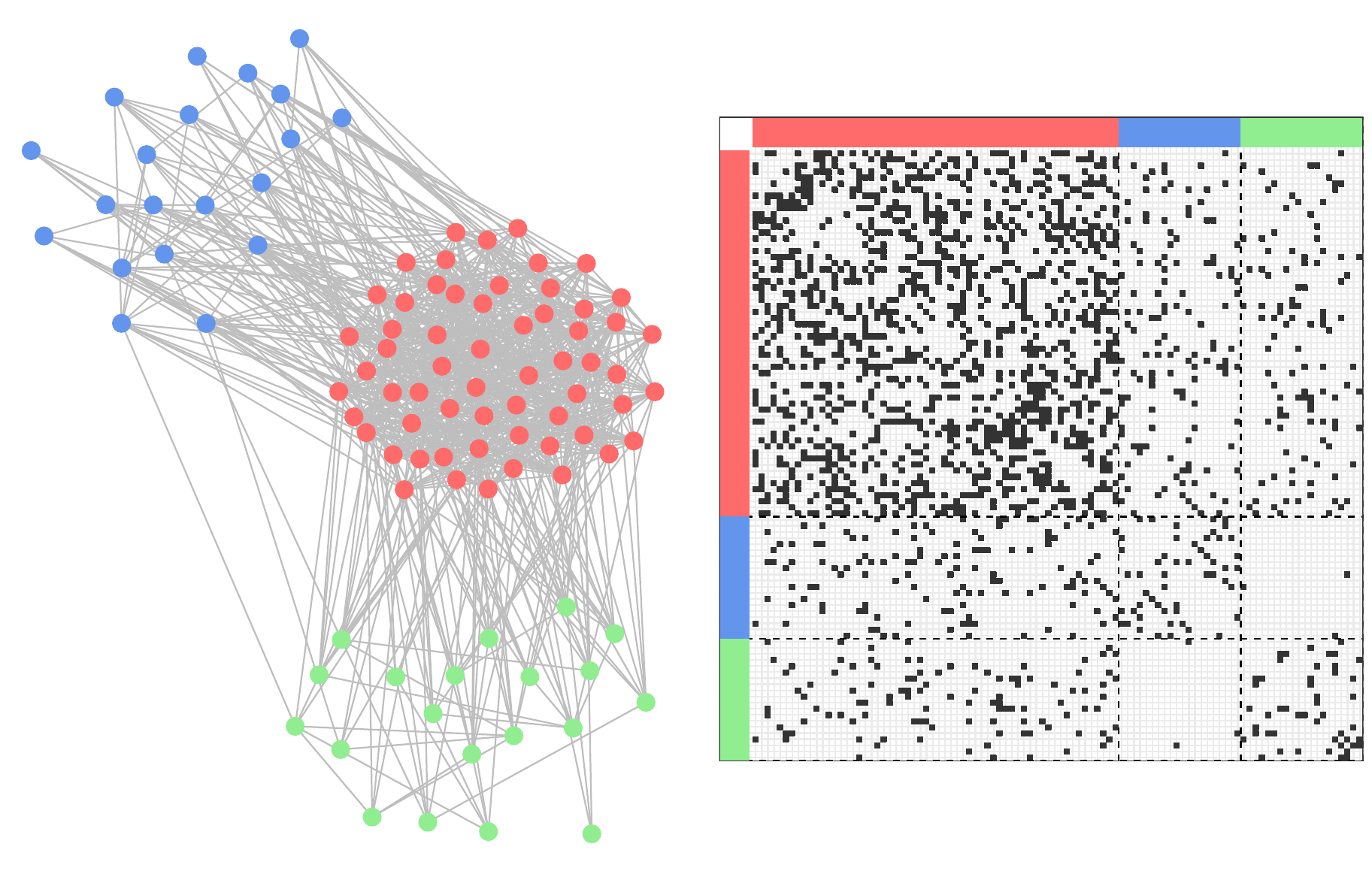}
    \caption{Right: synthetic network featuring a large and densely connected community (red) with links to peripheral nodes, which belong to two smaller communities (blue and green). Left: adjacency matrix with node memberships shown by the colored bar.} 
    \label{fig_star}
\end{figure}

The network, shown in the right panel of Figure \ref{fig_star}, consists of $3$ communities: a large community (the core) with 60 nodes and dense internal connections (shown in red), and two smaller communities (the peripheries), each with 20 nodes (shown in green and blue). The two peripheries are each sparsely connected internally and with limited connections between each other. Connections are generated according to the connectivity matrix:
\begin{equation}\label{eq:P}
  P = 
  \begin{pmatrix}
  0.30 & 0.085 & 0.085 \\
  0.085 & 0.13  & 0.01 \\
  0.085 & 0.01 & 0.13
  \end{pmatrix}.
\end{equation}
The adjacency matrix is shown in the left panel of Figure \ref{fig_star}. Detection of the two smaller communities is expected to be difficult because of both the limited number of connections within and between them and of their similar connectivity patterns with the central community. We discuss this in more details later.

We use Algorithms \ref{alg:SBM_k} and \ref{alg:SBM_k_ass} with $k=3$, initializing the node labels at random. For consistency, a uniform prior is used on $P_{ab}$ in the standard SBM, i.e. $\alpha=\beta=1$ in \eqref{eq:prior_Pab}. Also unit shape parameter $\gamma=1$ is used for the symmetric Dirichlet prior on assignment probabilities, cf. \eqref{eq:prior_pi}. 
We run each algorithm for 2000 iterations after a burn-in of 500. We repeat each run $100$ times to account for variability due to labels' initialization. For each run we estimate posterior community assignments based on 2000 posterior samples of label vector $z$. To this end, we use the R package \texttt{salso}, cf. \cite{Dah:Joh:Mul:22}, with default loss function (variation of information). Henceforth we refer to these estimates simply as posterior clusterings. 

\begin{figure}[ht]
    \centering
    \includegraphics[scale=.6]{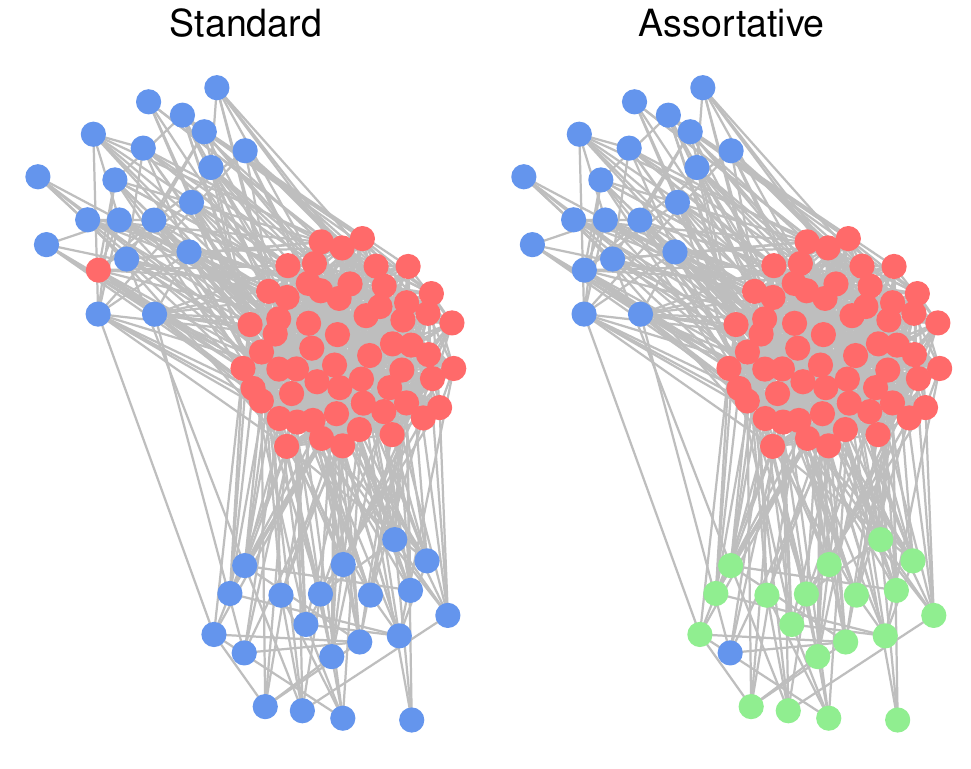}
    \caption{
    Posterior clustering of standard (left) and assortative (right) SBM.}
    \label{fig:synthetic}
\end{figure}

\begin{figure}[ht]
    \centering
    \includegraphics[scale=.6]{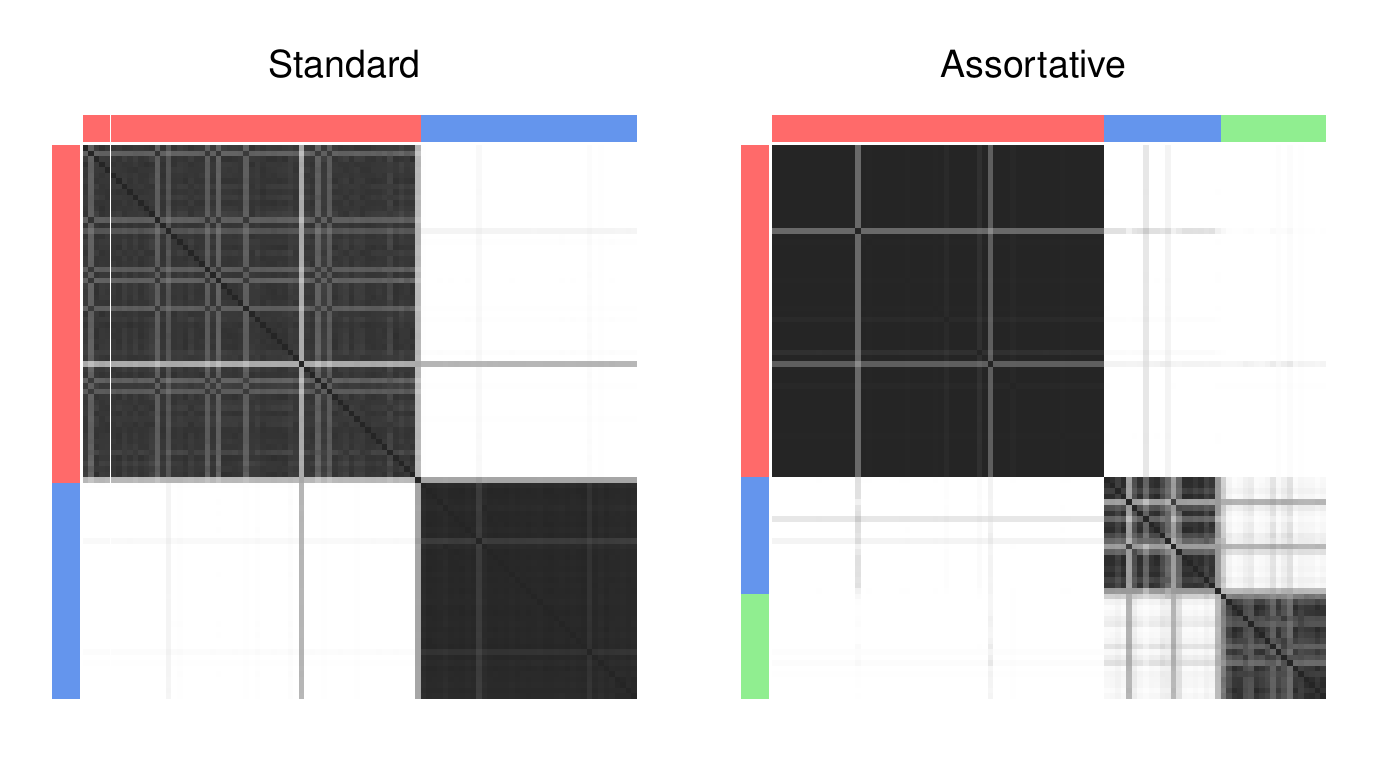}
    \caption{
     Posterior similarity matrix for standard (left) and assortative (right) SBM. 
}
    \label{psm}
\end{figure}

Out of the 100 runs, the assortative method successfully identified 3 communities every times, whereas the standard method never did. The posterior clusterings of a representative run are displayed in Figure \ref{fig:synthetic}. Standard SBM (left panel) identifies the central community but merges the peripheral nodes into a single second community. The nodes assigned to the third community vary between iterations, making it irrelevant for posterior clustering estimation. See the posterior similarity matrix in Figure \ref{psm}. In contrast, the assortative SBM (right panel) successfully identifies all 3 communities, particularly by distinguishing the two peripheries. 

To further investigate the differences between the standard and assortative methods, we focus on the representative run in Figure \ref{fig:synthetic} by examining the posterior distribution of the connectivity matrix $P$. Specifically, we consider a $\hat k\times \hat k$ matrix $P$, where $\hat k$ is the number of blocks in the estimated posterior clustering: $\hat k=2$ for the standard SBM and $\hat k=3$ for the assortative case. To account for label switching, we order the blocks by decreasing size across posterior samples. The posterior averages, denoted $\Pmodal$, are reported in \eqref{eq:double-matrix}.  The standard SBM violates the assortativity condition, which requires all diagonal elements to be larger than all off-diagonal elements. In contrast, the assortative SBM correctly recovers an assortative structure:
\begin{equation}
\label{eq:double-matrix}
 \text{standard }\Pmodal = 
  \begin{pmatrix}
  0.318 & 0.099 \\
  0.099 & 0.078 
  \end{pmatrix},
\quad \text{assortative }
\Pmodal = 
  \begin{pmatrix}
  0.325 & 0.098 & 0.094 \\
  0.098 & 0.134  & 0.016\\
  0.094 & 0.016 & 0.161
  \end{pmatrix}
\end{equation}
The posterior estimate of $P$ from the standard method is consistent with the empirical estimates based on the two-cluster solution obtained by merging the two peripheries: $\hat P_{22}=0.068$ compared to $\hat P_{12}=0.098$, which violates the assortativity constraint. 

\begin{figure}[ht]
    \centering
    \includegraphics[scale=.6]{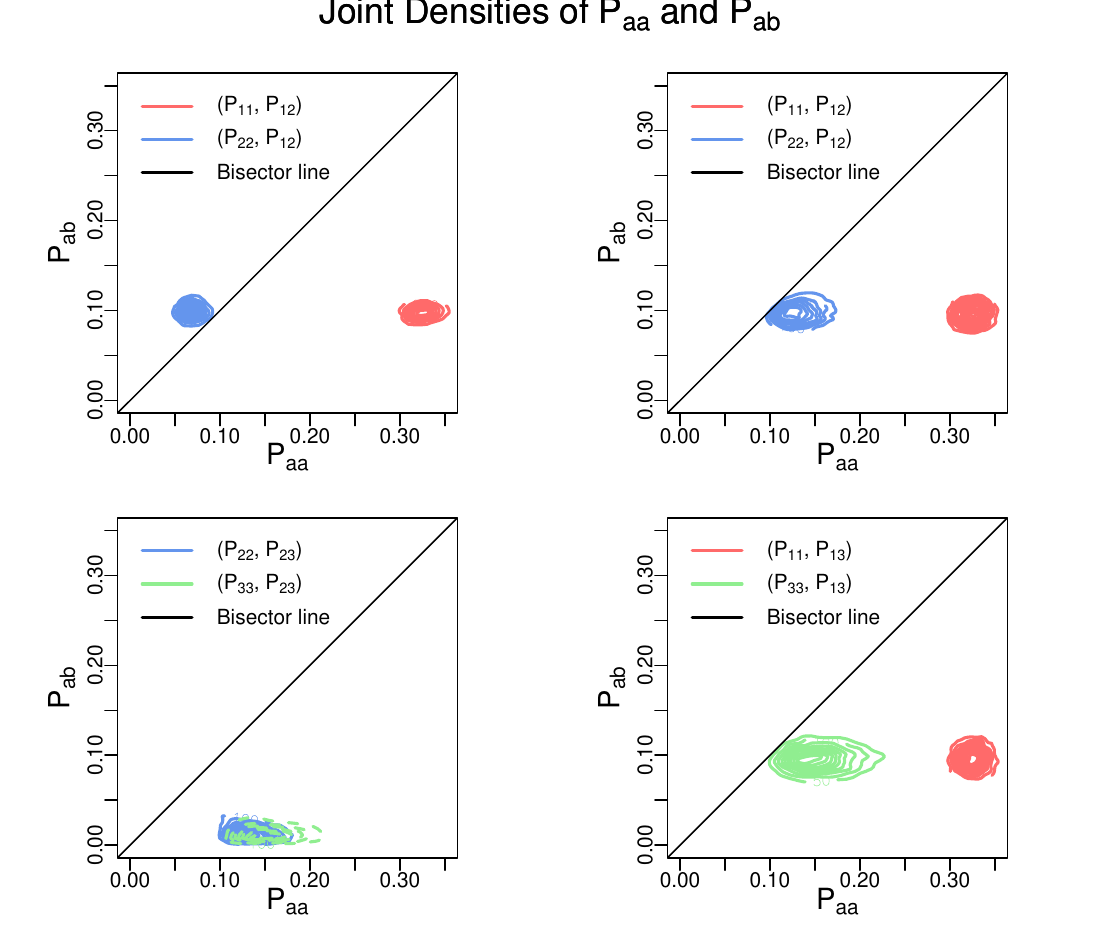}
    \caption{Joint density of between vs within probabilities for standard (top left) and assortative case (bottom left, top and bottom right). For each $a\neq b$ we plot the joint posterior densities $(P_{aa},P_{ab})$ and $(P_{bb},P_{ab})$.
    Coloring of classes is consistent with that of Figure \ref{fig:synthetic}.}
    \label{fig:posterior}
\end{figure}

Finally we report bivariate posterior contour plots of $P$ in Figure \ref{fig:posterior}, to be compared with the prior densities shown in Figure \ref{fig:prior}. 
In each panel, we compare a between-block probability $P_{ab}$ (vertical axis) with the corresponding two within-block probabilities, $P_{aa}$ and $P_{bb}$ (horizontal axis). In the assortative case, the bivariate plots involving $P_{12}$ and $P_{13}$ are nearly identical, reflecting the true connectivity matrix (cf. \eqref{eq:P}). For the bivariate plot involving $P_{23}$, the contour plots almost overlap due to the equal within-block probabilities of the two peripheral clusters (cf. again \eqref{eq:P}). As expected, the standard method produces a bivariate contour plot of $P_{12}$ against $P_{22}$ above the diagonal, contradicting the assortativity constraint. This cannot happen in the assortative case, since the prior assigns zero mass to that region. 

To put the above results into context, the SBM used to generate the network displayed in Figure \ref{fig_star} corresponds to a weak-recovery scenario as far as detection of the two peripheries is concerned. Consider the subnetwork consisting of these nodes, so $n=40$ in total, with within-block probability of $p=0.13$, and a between-block probability of $q=0.01$. The Kensten-Stigum bound is barely satisfied \citep{Dec:etal:11}, as the signal-to-noise ratio $n(p-q)^2/(2(p+q))=2.057$ is just above the critical threshold of $1$. 
We recall here that the Kensten-Stigum bound characterizes the conditions under which polynomial-time algorithms can find a partition correlated with a {\it planted partition} -- that is, equally sized groups with common within- and between-block probabilities. Therefore, when generating networks according to $P$ in \eqref{eq:P}, some realizations may have a signal too weak for either the standard or the assortative SBM to correctly identify the two peripheries. Conversely, for other realizations, the signal may be strong enough for both methods to recover these communities correctly. 
For intermediate level of signal strength, the connections of peripheral nodes with the central ones play a crucial role. Since $P_{13}$ and $P_{23}$ are equal, the peripheral nodes behave similarly in their connections to the core, causing the likelihood to favor the two-cluster solution. To illustrate this, consider the complete likelihood \eqref{eq:likelihood_sbm} of the two competing clusterings. 
Let $L_{\text{true}}=p(A,z_{\text{true}}|P,\hat\pi)$, where $z_{\text{true}}$ is the true label assignment, $P$ is given in \eqref{eq:P}, and $\hat\pi=(0.6,0.2,0.2)$ represents the relative sizes corresponding to $z_{\text{true}}$. This is to be contrasted with $L^{\star}=p(A,z^{\star}|P^{\star},\hat\pi^\star)$ where $z^{\star}$ is the label assignment corresponding to the two-cluster solution, $\hat\pi^\star=(0.6,0.4)$, and $P^\star$ is the corresponding  $2\times 2$ connectivity matrix 
  $$P^{\star} = 
  \begin{pmatrix}
  0.30 & 0.085 \\
  0.085 & 0.068 
  \end{pmatrix}.$$
Here $0.068$ is the average degree for a peripheral node.
It turns out that
  $\log L_{\text{true}}=-1791.792$
compared to
  $\log L^{\star}=-1800$
when using for $A$ the expected adjacency matrix, i.e., the node-wise connectivity matrix. When using the actual adjacency matrix as in Figure \ref{fig:synthetic}, the values become  $-2150.93$ and $-2151.107$, respectively. Thus, we observe two nearly equivalent local maxima in the likelihood, which are likely to appear as posterior modes in the standard SBM as well. Enforcing assortativity through the prior prevents the posterior from placing mass around the {\it unfavorable mode} corresponding to the two-cluster solution. Therefore, the assortative SBM has an advantage in cases of intermediate signal strength. 

To illustrate this, we conducted a more extensive simulation by generating $100$ networks from the SBM defined by \eqref{eq:P}. For each network, we ran both the standard and the assortative Gibbs samplers for $10$ different label initializations of the node labels. We focus on networks for which the posterior clustering was robust to initialization -- that is, it consistently produced either two or three clusters across all $10$ runs. 
The standard SBM estimated three clusters in $5$ cases and two clusters in $68$ cases. The assortative SBM estimated three communities in $49$ cases and two communities in $35$ cases. For precisely $18$ networks, the standard SBM identified two communities, while the assortative SBM identified three. The network shown in Figure \ref{fig_star} is representative of these $18$ cases.

In conclusion, incorporating assortativity into Bayesian community detection can yield significantly different clustering results. In the next section, we address the case where the number of communities $k$ is unknown.

\section{Unknown number of communities}\label{section:3}
A key challenge in community detection is the estimation of the number of groups \( k \). Bayesian inference is particularly well-suited for this task, as it offers a probabilistic framework for the joint inference of $k$, label assignments \( z \) and model parameters. By placing a prior distribution on $k$, one can construct an MCMC that updates $k$ at each iteration, along with $z$ and $P$. However this approach often requires complex MCMC moves to explore parameter spaces of varying dimension, such as reversible jump MCMC of \citet{green1995}, that is known to suffer from a lack of scalability and mixing issues. Depending on the model specification and prior choices, different marginalization strategies can be employed. For instance \citet{mcdaid:murphy:friel} considered Poisson-weighted edges, and proposed integrating out both the assignment probabilities $\pi$ and the block parameters, thereby extending the allocation sampler of \citet{nobile:fearnside} to perform MCMC sampling over $k$ and $z$.
This approach can also suffer from mixing issues due to the complexity of Metropolis moves that simultaneously update $(k, z)$. A related strategy is the microcanonical SBM formulation introduced by \citet{Pei:17}. The prior on the nodes partition is constructed in a non-informative way, which comes at the cost of loosing the ability to use prior predictive probabilities when devising Gibbs updates of $z$ (see below). Moreover, a maximum entropy prior is placed on block parameters, which induces an hard constraint on the number of connections between communities. 
An alternative approach that has recently experienced a surge of applications consists in integrating out $k$ and implementing Gibbs sampler over $z$ and $P$, and estimate $k$ through the number of group assignments featured by $z$, \citet{Gen:Bha:Pat:19}. 

We briefly recall the theoretical framework. In the setting of Section \ref{section:2}, we shift attention from the distribution of the assignment labels $z$ to the distribution of the partition of the $n$ nodes that these labels induce. This is most naturally described by the exchangeable partition probability function (EPPF), a symmetric function of the block sizes of the partitions, denoted  
$p_n(n_1,\ldots,n_{k_n})$, where $k_n\leq k$ is the number of blocks and $n_1,\ldots,n_{k_n}$ are their sizes in an arbitrary order. For $x_{(t)}=x(x-1)...(x-t+1)$ the rising factorial, one obtains from  \eqref{eq:p_z} the following expression:
\begin{equation}\label{eq:eppf}  
  p_n(n_1,\ldots,n_{k_n})=
  \frac{\prod_{i=1}^{k_n}(k\gamma-i\gamma)}{(k\gamma+1)_{n-1}}
  \prod_{i=1}^{k_n}(1+\gamma)_{n_i-1}.
\end{equation}
This can be written as the product of a nonnegative weight, say $V_{n,k_n}$, depending only on $k_n$ and $n$, and the term $\prod_{i=1}^{k_n}(1+\gamma)_{n_i-1}$. \citet{Gne:Pit:06} characterized all exchangeable random partitions with this product form, 
  $$p_n(n_1,\ldots,n_{k_n})=
  V_{n,k_n}
  \prod_{i=1}^{k_n}(1-\sigma)_{n_i-1}$$
in terms of a discount parameter $\sigma\in(-\infty,1)$ and the array of weights $V_{n,k}$ satisfing the recursion
  $$V_{n,k}=(n+\gamma k)V_{n+1,k}+V_{n+1,k+1}.$$ 
These are known as Gibbs-type partitions, and they exhibit two distictive large-$n$ behaviors for the number of blocks $k_n$: divergence to infinity for positive $\sigma$, and convergence to a possibly random limit for negative $\sigma$. \citet{Lij:Ram:Pru:07} were the first to construct random probability measures that induce Gibbs-type partitions with positive $\sigma$ and to derive the corresponding prediction rules. 
For the negative-$\sigma$ case, \citet{Gne:Pit:06} showed that the EPPF coincides with that of a symmetric Dirichlet distribution \eqref{eq:eppf} with a distribution on the number of blocks. \citet{Gne:10ecp} analyzed a special case where $k_n$ converges almost surely to a finite random variable but has infinite mean. \citet{Deb:Lij:Pru:13} systematically studied random probability measures -- thus going beyond the partitions-based framework of \citet{Gne:Pit:06} -- that induce Gibbs-type partitions with negative $\sigma$. As special cases, they also considered the Poisson and negative binomial distributions on the number of partition blocks and derived the corresponding prediction rules explicitly. With either the predictive distributions or the partition distributions in hand, a Gibbs sampler for mixture models based on these random probabilities follows directly by plugging them into any standard  Dirichlet process mixture sampler (cf. \cite{Nea:00}). A review of a whole stream of papers on Gibbs-type priors is provided in \citet{Deb:etal:15}. \citet{mill:harr:2018} later referred to the negative-$\sigma$ Gibbs-type setting in mixture-modeling applications as {\it mixture of finite mixtures}.

Let us focus to the negative-$\sigma$ case, namely $\sigma=-\gamma$ for $\gamma$ the scale parameter of the symmetric Dirichlet prior in \eqref{eq:prior_pi}. Given a prior $p(k)$ on $k$ supported on $\{1,2,\ldots\}$, one gets the the weight $V_{n,k}$ upon marginalization
\begin{equation} \label{eq:V_kn} 
      V_{n,k}=\sum_{m\geq k} \frac{\prod_{i=1}^{k-1}(m-i)\gamma}{(m\gamma+1)_{n-1}}p(m),
\end{equation}
and, in turn, the prior predictive probability
\begin{equation}\label{eq:prior_classi_a}
  \bbP(z_i=a| z_{-i})=\frac{V_{n,k_{-i}}}{V_{n-1,k_{-i}}}(n_a( z_{-i})+\gamma),
  \quad n_a( z_{-i})>0,
\end{equation}
that is the probability of allocating node $i$ to an existing block labelled $a$ among the $k_{-i}$ featured by the $n-1$ assignments in $z_{-i}$. Cf. \eqref{eq:prior_predictive}. The prior predictive probability of allocating element $i$ to a new block is also recovered as
\begin{equation}\label{eq:prior_classi_new}
  \bbP(z_i=\text{new}| z_{-i})
  =\frac{V_{n,k_{-i}+1}}{V_{n-1,k_{-i}}}.
\end{equation}
\citet{Gen:Bha:Pat:19} considered a truncated Poisson prior on $k$, for which \eqref{eq:V_kn} does not have a closed-form solution. We consider instead the distribution introduced by \citet{Gne:10ecp}, namely
\begin{equation}\label{eq:Gnedin}
 k\sim p(k)=\frac{\lambda(1-\lambda)_{(k-1)}}{k!}\Indic_{\{1,2,\dots\}}(k)
\end{equation}
that, when paired with unit scale $\gamma=1$ Dirichlet parameter 
yields weight $V_{n,k}$ in closed form:
\begin{equation}\label{vn-explicit}
  V_{n,k}=
  \frac{(k-1)!(1-\lambda)_{k-1}(\lambda)_{n-k}}{(n-1)!(1+\lambda)_{n-1}}, 
  \quad \lambda \in (0,1),  
\end{equation}
cf \citet[Sect 4.1]{Deb:Lij:Pru:13}. By exploiting the conjugate prior \eqref{eq:prior_Pab} for standard SBM, \citet{Gen:Bha:Pat:19} adapted Algorithm 2 in \citet{Nea:00} to update labels $z$ and $P$. See also \citet[Example 4]{Leg:Rig:Dur:Dun:22} for a slightly different version of the sampler that marginalizes $P$ in accordance to Algorithm 3 in \citet{Nea:00}. It is important to note that such marginalization heavily relies on the beta-binomial conjugacy recalled in \eqref{eq:full_Pab}. 

We discuss next the algorithm of \citet{Gen:Bha:Pat:19} upon the adoption of Gnedin prior \eqref{eq:Gnedin}. Specifically, one finds the following expressions for the prior predictive probabilities \eqref{eq:prior_classi_a} and \eqref{eq:prior_classi_new}:
\begin{align*}
    \bbP(z_i=a| z_{-i})
    &=\begin{cases}
    \displaystyle
    \frac{n-1-k_{-i}+\lambda}{(n-1)(n-1+\lambda)}(n_a(z_{-i})+1)
    &\text{if $a$ is an existing cluster}\\
    \displaystyle
    \frac{k_{-i}(k_{-i}-\lambda)}{(n-1)(n-1+\lambda)}
    &\text{if $a$ is a new cluster}
    \end{cases}
\end{align*}
where \comillas{existing} or \comillas{new cluster} are to be intended with respect to $z_{-i}$. Recall equation \eqref{eq:like_contr} of Section \ref{section:2} for 
the likelihood contribution to $z_i=a$. When $a$ is a new cluster, the probabilities $P_{a\,z_j}$ need to be marginalized out from this expression with respect to the prior \eqref{eq:prior_Pab}. One obtains
\begin{equation}\label{eq:m(A_i)}
    m(A_i):=
    \prod_{b=1}^{k_{-i}}
    \text{Beta}\big(
    \sum\nolimits_{j\neq i, z_j=b}A_{ij}
    +\alpha
    ,n_{b}(z_{-i})-
    \sum\nolimits_{j\neq i, z_j=b}A_{ij}
    +\beta\big)
\end{equation}
where $\text{Beta}(\cdot,\cdot)$ is the beta integral, $\sum_{j\neq i, z_j=b}A_{ij}$ is the number of edges between node $i$, newly assigned to a group on its own, and nodes in group $b$, and $n_{b}(z_{-i})$ is the maximum possible number of edges between these two groups. 
  $
  $

With these notations at hand we can express the full conditionals of $z$ as follows:
\begin{align}
  P(z_i=a| z_{-i}, A, P)
  \label{eq:full_z_npmic}
  &\propto
  \begin{cases}
  \displaystyle
    (n_a(z_{-i})+1)
    \prod_{j\neq i} P_{a\,z_j}^{A_{ij}}
  (1-P_{a\,z_j})^{(1-A_{ij})}, \ \ 
    &\text{$a$ existing cluster}\\
  \displaystyle
    \frac{k_{-i}(k_{-i}-\lambda)}
    {n-1-k_{-i}+\lambda} m(A_i), \ \ 
    &\text{$a$ new cluster.}
  \end{cases}
\end{align}
If the new value of $z_i$ is not associated with any other node, draw values of $P_{aa}$ and $P_{ab}$ from the prior in \eqref{eq:prior_Pab}. Note that if node $i$ previously belonged to a singleton group, some bookkeeping is required before updating the next node. In particular, the connectivity matrix $P$ must be reshaped by removing the row and the column corresponding to the now-empty community. We report the Gibbs sampling algorithm next.

\begin{algorithm}[ht]
\caption{Gibbs sampler for SBM, $k$ unknown}\label{alg:geng}
\begin{algorithmic}
\Procedure{SBM}{}
\State{Require $n\times n$ adjacency matrix $A$, number of iterations $M$,} 
\State{prior hyperparameters $\alpha,\beta,\lambda$.}
\smallskip
\State {Initialize $z=(z_1,...,z_n)$ 
}
\smallskip
\For{each iter $j=1$ to $M$} 
    \State{- update $ P=(P_{ab})$ conditional on $z$ and $A$ from \eqref{eq:full_Pab} } 
    \For{each iter $i=1$ to n} 
    \State{- update $z_i$ conditional on $z_{-i}$, $P$ and $A$ from \eqref{eq:full_z_npmic} via  \eqref{eq:m(A_i)}
    }
    \EndFor
\EndFor
\EndProcedure
\end{algorithmic}
\end{algorithm}


In the next section we adapt the sampler to the case of assortativity-constrained probability matrix $P$.


\subsection{Assortative SBM - k unknown}\label{subsection:proposal}


Let now the prior on $P$ be specified according to \eqref{eq:prior_P_ass}. For the sake of completeness, we report the full prior specification next, where the dependence on $k$ is highlighted:
\begin{align*}    
    k\sim p(k) & = \frac{\lambda(1-\lambda)_{(k-1)}}{k!}\Indic_{\{1,2,\dots\}}(k),\\
    \pi|k &\sim\text{Dirichlet}(1,\ldots,1),\\ 
    \epsilon &\sim U(0,1),\\
    P_{aa}|\epsilon,k & \overset{\text{iid}} \sim  \text{U}(\epsilon,1), \quad  0\leq a\leq k,\\
    P_{ab}|\epsilon,k & \overset{\text{iid}} \sim  \text{U}(0,\epsilon), \quad  0\leq a < b\leq k.
\end{align*}
The main point to emphasize here is that the loss of the beta-binomial conjugacy affects not only the update of the probability matrix $ P$, but it also prevents the evaluation in closed-form of the {\it marginal} likelihood contribution $m(A_i)$, cf. \eqref{eq:m(A_i)}. More in details, upon defining the incomplete beta integral $\text{Beta}_{(0,\ \epsilon)}(\alpha,\beta)=\int_0^\epsilon u^{\alpha-1}(1-u)^{\beta-1}\ddr u$, we have
\begin{equation*}
    m(A_i):=
    \prod_{b=1}^{k_{-i}}
    \text{Beta}_{(0,\ \epsilon)}\big(
    \sum\nolimits_{j\neq i, z_j=b}A_{ij}
    +1
    ,n_{b}(z_{-i})-
    \sum\nolimits_{j\neq i, z_j=b}A_{ij}
    +1\big).
\end{equation*}
However, the incomplete beta integral can become computationally expensive, and the accuracy of numerical approximation critical in the evaluation of the probability of assignment of a node to a new cluster. Hence we adopt the auxiliary variable strategy proposed in Algorithm 8 of \citet{Nea:00} to avoid direct evaluation of the integral.
%
%
Specifically, we introduce $m\geq 1$ auxiliary communities to deal with the assignment of a node to a new cluster in the absence of conjugacy. Let $k$ denote the number of clusters featured by $z$, and let $k_{-i}$ be the number of clusters among the $n-1$ assignments in $z_{-i}$. Note that $k_{-i}=k-1$ or $k$ if node $i$ forms a singleton cluster, and $k_{-i}=k$ otherwise. We label the $m$ candidate new clusters as $k+1,...,k+m$, and draw the corresponding edge probabilities $P_{ab}$ --  $(k+1)+\cdots+(k+m)$ in total -- from the prior \eqref{eq:prior_P_ass}, conditional on the current value of the cut-off parameter $\epsilon$. The intractable marginal likelihood contribution $m(A_i)$ is then replaced by the {\it Monte Carlo average}
  $$\frac1m \sum_{a=k+1}^{k+m}
  \prod_{j\neq i} P_{a\,z_j}^{A_{ij}}
  (1-P_{a\,z_j})^{(1-A_{ij})}$$
and so it enters the probability that $z_i$ is assigned to a new cluster labeled $a$ accordingly:
\begin{equation}
  \label{eq:full_z_npmic_ass}
  P(z_i=a|z_{-i}, A, P)
  \propto
    \begin{cases}
    \displaystyle
    (n_a(z_{-i})+1)
    \prod_{j\neq i} P_{a\,z_j}^{A_{ij}}
    (1-P_{a\,z_j})^{(1-A_{ij})}, \ \ 
    &\text{if $a$ is an existing cluster}\\
    \displaystyle  
    \frac{k_{-i}(k_{-i}-\lambda)}{n-1-k_{-i}+\lambda}\frac{1}{m}
    \prod_{j\neq i} P_{a\,z_j}^{A_{ij}}
  (1-P_{a\,z_j})^{(1-A_{ij})},  
  &\text{for $a=k+1,\ldots, k+m$.}
        \end{cases}
\end{equation}
Proper bookkeeping is required before proceeding to the update of the next node; this is even more crucial than in Algorithm \ref{alg:geng}, due to the need to discard temporary auxiliary communities. 


The update of the probabilities in $P$ has been described in Section \ref{section:2}, see equations \eqref{eq:full_Pab_ass} and \eqref{eq:full_epsilon} therein. As for the cutoff variable $\epsilon$, it is important to generate a value even when $k=1$, since this value will be needed in the subsequent update of $z_i$ in view of the creation of the $m$ auxiliary communities. Below, we detail the procedure for generating $\epsilon$ using auxiliary variables, cf. Section \ref{section:2}:
\begin{equation*}
\begin{split}
  &
  y|\epsilon \sim U(0,(1-\epsilon)^{-k}),\quad  
  w \sim U(0,1)  \\
  \epsilon &=
  \begin{cases}
1-(1-P_{11})^w, &k=1\\
\exp(w(\log p-\log \bar q)+\log \bar q), &k=2\\
(w(p^{1-k(k-1)/2}-{\bar q}^{1-k(k-1)/2})+{\bar q}^{1-k(k-1)/2})^{2/(2-k(k-1))}, &k>2\\
\end{cases}
  \end{split}
\end{equation*}
with $p=\min_{a}P_{aa}$, $q=\max_{a< b}P_{ab}$, and $\bar q=\max\{q,1-y^{-1/k}\}$. We report the Gibbs sampler algorithm next. 

\begin{algorithm}ht]
\caption{A Gibbs sampler for assortative-SBM, $k$ unknown}\label{alg:adb}
\begin{algorithmic}
\Procedure{a-SBM}{}
\State{Require $n\times n$ adjacency matrix $A$, number of iterations $M$,} 
\State{prior hyperparameters $\alpha,\beta,\lambda$ and $m$.}
\smallskip
\State {Initialize $z=(z_1,...,z_n)$ and $\epsilon$
}
\smallskip
\For{each iter $j=1$ to $M$} 
    \State{- update $ P=(P_{ab})$ conditional on $\epsilon$, $z$ and $A$ from \eqref{eq:full_Pab_ass} } 
    \State{- update $\epsilon$ conditional on $P$, $z$ and $A$ from \eqref{eq:full_epsilon} } 
    \For{each iter $i=1$ to n} 
    \State{- update $z_i$ conditional on $z_{-i}$, $P$ and $A$ from \eqref{eq:full_z_npmic_ass}
    }
    \EndFor
\EndFor
\EndProcedure
\end{algorithmic}
\end{algorithm}




\section{Simulation study}\label{section:simulation}
In Section \ref{subsection:motivating-example-section} we considered a synthetic network generated from a SBM where the inclusion of the assortativity constraint through the prior significantly improved community detection. Here, we further investigate the benefits of assortativity when the number of communities $k$ is unknown. To this end, we conduct an extensive simulation study generating networks according to  \citet{Lan:For:Rad:08}, which provides state of the art benchmarks for evaluating community detection algorithms. Networks are generated so to reflect realistic features that the SBM fails to capture, such as flexible degree and community size distributions. Each network is generated with parameters including the number of nodes $n$, minimum and maximum community sizes $n_{min}$ and $n_{max}$, average degree $\langle d\rangle$, maximum degree $d_{max}$, mixing parameter  $\mu$ (ratio of out-group to total connections), and power-law exponents $t_1$ and $t_2$ controlling degree and community size distributions, respectively. We generated $N=100$ networks with $n=200$, $t_1=t_2=2$, $n_{min}=5$, $n_{max}=50$, $d_{max}=49$, and for each combination of mixing parameter $\mu \in \{0.1, \dots, 0.6\}$ and mean degree $\langle d\rangle\in \{10,15,20,25\}$, $24$ combinations in total. 
These choices follow the guidelines of \citet{Lan:For:Rad:08} to ensure realistic and challenging conditions for community detection. 

Each network is analyzed by two runs of the Algorithm \ref{alg:adb} for assortative SBM and two runs of Algorithm \ref{alg:geng} for standard SBM. Prior hyperparameters are set as in Section \ref{subsection:motivating-example-section}, that is $\alpha=\beta=1$ in \eqref{eq:prior_Pab},   together with $\lambda=0.45$ in \eqref{eq:Gnedin}. The latter corresponds to a prior distribution on the number of communities featured by the network of $n=200$ nodes fairly flat, centered around $16$ with standard deviation around $40$, cf. 
\citet[Section 4.2]{San:Fri:25}.
Consider that the synthetic networks feature a number of communities $k$ ranging from a minimum of $4$ to a maximum of $40$. The algorithms are referred to as \comillas{a-sbm} and \comillas{sbm}, respectively, in plots and tables below. Different random label initializations are used in each run to ensure robustness. Each run generates $3000$ posterior samples after a burn-in of $1000$ iterations. Thinning of size $5$ results in two sets of $600$ posterior sample, henceforth referred to as an individual {\it run}. They are used to compute the potential scale reduction factor $\hat{R}_{D}$ \citep[Chapter 11]{gelman} for the parameter deviance 
\begin{equation}\label{eq:deviance} 
   D:=-2 \sum_{i<j}    \log \biggl(\sum_a \sum_b P_{ab}^{A_{ij}}(1-P_{ab})^{(1-A_{ij})}\frac{n_an_b}{n^2}\biggr).
\end{equation}
Convergence is assessed using the standard practical threshold $\hat{R}_{D} < 1.1$.  

\begin{figure}[htbp]
    \centering
    \includegraphics[width=0.9\linewidth]{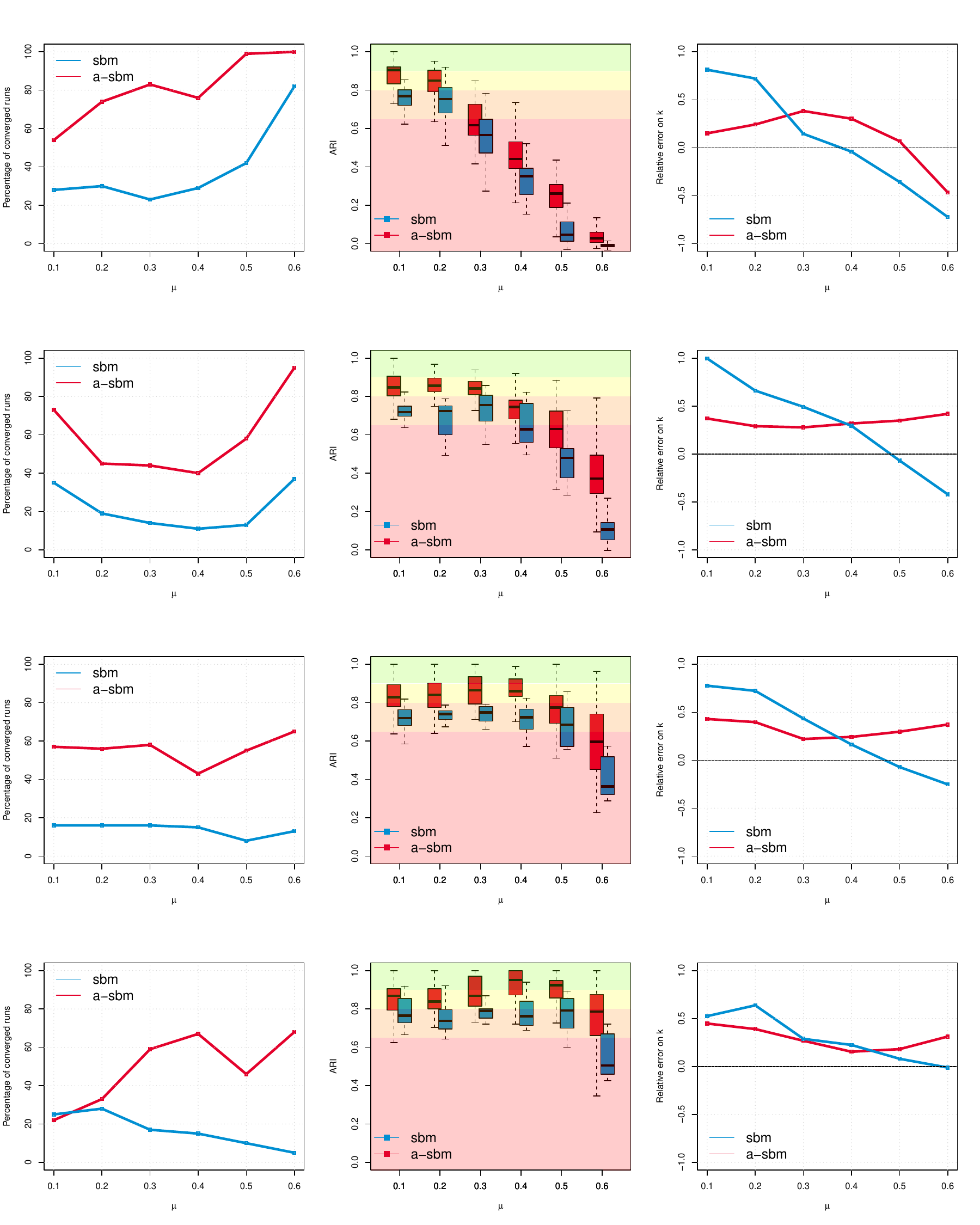}
    \caption{Networks of size $n=200$ generated according to \citet{Lan:For:Rad:08}. Rows display average degree in increasing order, $\langle d \rangle\in\{10,15,20,25$\}.
    Left: number of runs that achieved convergence $\hat{R_D} < 1.1$. Center: adjusted Rand index for runs that achieved convergence. The background colors reflect the recovery level of ARI so to help interpretation. 
    Right: relative error of estimation of $k$ over runs that reached convergence.}
    \label{fig:s1}
\end{figure}

The results are summarized in Figure \ref{fig:s1}. 
Each row corresponds to a different average degree $\langle d \rangle$ arranged in increasing order top-down. The horizontal axis in each plot represents the mixing parameter $\mu$. The left panel displays the percentage of runs that achieved convergence ($\hat{R}_{D} < 1.1$). For these converged runs, the middle panel shows the adjusted Rand index (ARI) comparing the posterior clustering to the true one, while the right panel presents the average of the relative error in estimating the number of communities $k$, that is $(\hat  k-k_{\text{true}})/k_{\text{true}}$ averaged over runs that reached convergence. 


As expected, the higher the average degree $\langle d\rangle$, the stronger the signal in the network, yielding a more accurate community detection. Performances also vary with the mixing parameter $\mu$; in particular, it deteriorates as $\mu$ increases, reflecting less assortative communities. Notably, the assortative SBM consistently outperforms standard SBM in terms of ARI. 
An interesting trend is observed in the recovery of the number of communities $k$. Assortative SBM overestimates $k$ fairly consistently across scenario except in the least informative ones, that is for low $\langle d \rangle$ and high $\mu$, as reflected by poor community detection (ARI below $0.2$). Standard SBM instead has a downward tendency with respect to the mixing parameter $\mu$, that is as assortativity fades out: it tends to overestimate $k$ under strong to moderate assortativity, underestimate $k$ for low assortativity. 
Since model misspecification generally results in an overestimation of $k$, the assortative SBM demonstrates a clear advantage over the standard SBM, yielding more accurate estimates of the true number of communities in low to intermediate recovery regimes.
This better performance is in accordance with the discussion in Section~\ref{subsection:motivating-example-section}.

The number of runs for which the samplers achieved convergence ($\hat{R}_D<1.1$) also varies with $\mu$, $\langle d \rangle$ and across methods. 
The plots in the left panel in Figure \ref{fig:s1} show an overall better performance of the assortative SBM compared to the standard SBM across all values of $\langle d \rangle$ and $\mu$. This is confirmed by higher effective sample sizes (ESS) as reported in Table \ref{tab:ess-nonfiltrata}. A better convergence can be explained in terms of the reduced space the posterior has to explore in the assortative case. 
Recall that a potential scale reduction  $\hat{R}_{D}$ of $1.1$ or below  indicates that the sampler explores the same posterior modes in the space of parameters and nodes' partitions regardless of the initialization. As discussed in Section \ref{subsection:motivating-example-section}, the assortative prior acts by reducing the parameter space, and so excluding regions corresponding to potential local posterior modes in which the standard SBM could get trapped. Further work is needed to clarify whether there is a pattern with respect to $\mu$ and $\langle d \rangle$.

For completeness, we report in Table \ref{tab:ess-filtrata} the ESS for runs that reached convergence. As expected, these values are higher than in Table \ref{tab:ess-nonfiltrata}, which further justifies basing our conclusions for community detection on these runs (cf. centre and right panels of Figure \ref{fig:s1}). That ESS remains somewhat low even after accounting for assessment of convergence should not be of concern. The samplers must cope with posterior distributions that support a large number of communities 
$k$, as is the case for networks generated under \citet{Lan:For:Rad:08}. Standard Gibbs steps updating one item at a time are well known to suffer from poor mixing in such settings: several Gibbs scans are necessary to traverse the space of partitions. The problem worsens as the number of blocks of the partition increases. In practical applications, one would typically run longer chains with potentially higher thinning 
that facilitate exploration of posterior modes.


\begin{table}[ht]
\caption{Mean and standard deviation of Effective Sample Size (ESS) per number of samples. Results refer to the deviance \eqref{eq:deviance} for combinations of $\langle d\rangle$ and $\mu$, under sbm and a-sbm.\label{tab:ess-nonfiltrata}}
\centering
\begin{tabular}{lllrrrrrr}

\hline
$\langle d\rangle$ & Model & ESS$_D/\text{Iter}$ & $\mu=1$ & $\mu=2$ & $\mu=3$ & $\mu=4$ & $\mu=5$ & $\mu=6$ \\
\hline
\multirow{4}{*}{10} 
  & \multirow{2}{*}{a-sbm}
      & mean & 0.1508 & 0.1409 & 0.2613 & 0.2609 & 0.3963 & 0.4635 \\
  &   & sd   & 0.2231 & 0.1283 & 0.2080 & 0.2394 & 0.2348 & 0.1983 \\
  & \multirow{2}{*}{sbm}
      & mean & 0.1783 & 0.1686 & 0.1095 & 0.0949 & 0.1087 & 0.4476 \\
  &   & sd   & 0.3303 & 0.3080 & 0.2308 & 0.1908 & 0.1644 & 0.3449 \\
\hline
\multirow{4}{*}{15} 
  & \multirow{2}{*}{a-sbm}
      & mean & 0.1891 & 0.2431 & 0.1348 & 0.0926 & 0.2277 & 0.4255 \\
  &   & sd   & 0.2187 & 0.3143 & 0.2136 & 0.1242 & 0.2601 & 0.2284 \\
  & \multirow{2}{*}{sbm}
      & mean & 0.2072 & 0.0888 & 0.0617 & 0.0655 & 0.0543 & 0.0944 \\
  &   & sd   & 0.3229 & 0.2047 & 0.1439 & 0.2015 & 0.1567 & 0.1651 \\
\hline
\multirow{4}{*}{20} 
  & \multirow{2}{*}{a-sbm}
      & mean & 0.1392 & 0.1867 & 0.3965 & 0.2454 & 0.3395 & 0.3201 \\
  &   & sd   & 0.2010 & 0.2367 & 0.4003 & 0.3528 & 0.3637 & 0.3301 \\
  & \multirow{2}{*}{sbm}
      & mean & 0.0635 & 0.0913 & 0.0847 & 0.0791 & 0.0581 & 0.0494 \\
  &   & sd   & 0.1488 & 0.2241 & 0.2317 & 0.2291 & 0.2051 & 0.1486 \\
\hline
\multirow{4}{*}{25} 
  & \multirow{2}{*}{a-sbm}
      & mean & 0.0846 & 0.1999 & 0.2394 & 0.3722 & 0.1741 & 0.4437 \\
  &   & sd   & 0.1945 & 0.3221 & 0.2839 & 0.3288 & 0.1941 & 0.3735 \\
  & \multirow{2}{*}{sbm}
      & mean & 0.1534 & 0.1570 & 0.1066 & 0.0978 & 0.0664 & 0.0212 \\
  &   & sd   & 0.3044 & 0.3109 & 0.2739 & 0.2549 & 0.2158 & 0.1005 \\
\hline
\end{tabular}

\end{table}

\begin{table}[ht]
\centering
\caption{Mean and standard deviation over runs that reached convergence of Effective Sample Size (ESS) per number of samples. Results refer to the deviance \eqref{eq:deviance} for combinations of $\langle d\rangle$ and $\mu$, under sbm and a-sbm.\label{tab:ess-filtrata}}
\begin{tabular}{lllrrrrrr}
\hline
$\langle d\rangle$ & Model & ESS$_D/\text{Iter}$ & $\mu=1$ & $\mu=2$ & $\mu=3$ & $\mu=4$ & $\mu=5$ & $\mu=6$ \\
\hline
\multirow{4}{*}{10} 
  & \multirow{2}{*}{a-sbm}
      & mean & 0.2582 & 0.1772 & 0.3065 & 0.3294 & 0.3997 & 0.4635 \\
  &   & sd   & 0.2593 & 0.1305 & 0.1997 & 0.2359 & 0.2334 & 0.1983 \\
  & \multirow{2}{*}{sbm}
      & mean & 0.5931 & 0.5282 & 0.4009 & 0.2842 & 0.2236 & 0.5383 \\
  &   & sd   & 0.3888 & 0.3622 & 0.3465 & 0.2750 & 0.2033 & 0.3145 \\
\hline
\multirow{4}{*}{15} 
  & \multirow{2}{*}{a-sbm}
      & mean & 0.2493 & 0.5142 & 0.2872 & 0.1980 & 0.3654 & 0.4455 \\
  &   & sd   & 0.2280 & 0.2911 & 0.2498 & 0.1388 & 0.2659 & 0.2165 \\
  & \multirow{2}{*}{sbm}
      & mean & 0.5495 & 0.4148 & 0.3501 & 0.4940 & 0.3395 & 0.2236 \\
  &   & sd   & 0.3407 & 0.3006 & 0.2244 & 0.4127 & 0.3143 & 0.2164 \\
\hline
\multirow{4}{*}{20} 
  & \multirow{2}{*}{a-sbm}
      & mean & 0.2277 & 0.3188 & 0.6657 & 0.5434 & 0.5979 & 0.4721 \\
  &   & sd   & 0.2292 & 0.2454 & 0.3197 & 0.3650 & 0.3012 & 0.3181 \\
  & \multirow{2}{*}{sbm}
      & mean & 0.3068 & 0.5085 & 0.4815 & 0.4915 & 0.6451 & 0.3243 \\
  &   & sd   & 0.2602 & 0.3288 & 0.3919 & 0.3949 & 0.4057 & 0.2947 \\
\hline
\multirow{4}{*}{25} 
  & \multirow{2}{*}{a-sbm}
      & mean & 0.3224 & 0.5684 & 0.3896 & 0.5431 & 0.3477 & 0.6438 \\
  &   & sd   & 0.3181 & 0.3324 & 0.2851 & 0.2672 & 0.1573 & 0.2807 \\
  & \multirow{2}{*}{sbm}
      & mean & 0.5669 & 0.5334 & 0.5960 & 0.6211 & 0.6159 & 0.3145 \\
  &   & sd   & 0.3782 & 0.3866 & 0.3963 & 0.3370 & 0.3720 & 0.3665 \\
\hline
\end{tabular} 
\end{table}




\section{Discussion}\label{section:discussion}
In this paper, we investigated the role of assortativity in Bayesian community detection based on stochastic block modeling. We focus on the case where the number of communities is unknown and estimated together with nodes' partition. We introduce an hard constraint by placing a prior on the edge probabilities which enforces stronger connectivity within than between communities. Despite the loss of conjugacy and associated analytical tractability, we develop an efficient marginal Gibbs sampler for posterior inference. Through a motivating example, we argued that enforcing assortativity helps identify communities in weak recovery regimes by steering the posterior away from clustering solutions that violate the assortativity constraint. 
An extensive simulation study based on benchmark networks further confirmed that the proposed method improves community detection accuracy, particularly in challenging settings characterized by low signals and heterogeneity in node degree and community sizes.


We intend to extend the proposed estimation procedure to accommodate {\it weak assortativity} \eqref{eq:weak_ass}, where the edge probability within a community is constrained to be greater than the probabilities between that community and others. This can be achieved by introducing a cut off variable for each community 
$$P_{bb}\sim\mathrm{Unif}(\epsilon_b,1),\quad
  P_{ab}\sim\mathrm{Unif}(0,\epsilon_a\wedge\epsilon_b)\quad a<b$$
compared to (strong)  $$P_{bb}\sim\mathrm{Unif}(\epsilon,1),\quad
  P_{ab}\sim\mathrm{Unif}(0,\epsilon)\quad a<b$$
and appropriately modifying the Gibbs sampler.
As discussed in \citet{Jia:Tok:21}, weak assortativity 
offers a minimal yet effective modeling condition for recovering the node partition through accurate estimation of the node-wise connectivity matrix. While theoretical results for community detection remains limited, see \citet{Gao:Vaa:Zho:20,Gen:Bha:Pat:19}, investigating posterior asymptotics to support our empirical findings represents a promising direction for future work. We also aim to speed-up mixing and improve computational efficiency of the marginal Gibbs sampler by exploring batch updates rather than sequential update of label assignments, in line with ideas put forward by \citet{Zha:Zho:20}.


\bibliographystyle{apalike}
\bibliography{biblio}

\end{document}